\documentclass[twocolumn, secnumarabic,amssymb, nobibnotes, aps, prl, floats, floatfix, superscriptaddress, amsmath]{revtex4}
\usepackage{graphicx}

\def\be{\begin{equation}} \def\ee{\end{equation}}

\newcommand{\ket}[1]{\mbox{$|#1\rangle$}}

\begin{document}

\title{Reduced phase error through optimized control of a superconducting qubit}
\author{Erik Lucero}
\affiliation{Department of Physics, University of California at Santa Barbara, Broida Hall, Santa Barbara, CA 93106}
\author{Julian Kelly}
\affiliation{Department of Physics, University of California at Santa Barbara, Broida Hall, Santa Barbara, CA 93106}
\author{Radoslaw C. Bialczak}
\affiliation{Department of Physics, University of California at Santa Barbara, Broida Hall, Santa Barbara, CA 93106}
\author{Mike Lenander}
\affiliation{Department of Physics, University of California at Santa Barbara, Broida Hall, Santa Barbara, CA 93106}
\author{Matteo Mariantoni}
\affiliation{Department of Physics, University of California at Santa Barbara, Broida Hall, Santa Barbara, CA 93106}
\author{Matthew Neeley}
\affiliation{Department of Physics, University of California at Santa Barbara, Broida Hall, Santa Barbara, CA 93106}
\author{A. D. O'Connell}
\affiliation{Department of Physics, University of California at Santa Barbara, Broida Hall, Santa Barbara, CA 93106}
\author{Daniel Sank}
\affiliation{Department of Physics, University of California at Santa Barbara, Broida Hall, Santa Barbara, CA 93106}
\author{H. Wang}
\affiliation{Department of Physics, University of California at Santa Barbara, Broida Hall, Santa Barbara, CA 93106}
\author{Martin Weides}
\affiliation{Department of Physics, University of California at Santa Barbara, Broida Hall, Santa Barbara, CA 93106}
\author{James Wenner}
\affiliation{Department of Physics, University of California at Santa Barbara, Broida Hall, Santa Barbara, CA 93106}
\author{Tsuyoshi Yamamoto}
\affiliation{Department of Physics, University of California at Santa Barbara, Broida Hall, Santa Barbara, CA 93106}
\affiliation{Green Innovation Research Laboratories, NEC Corporation, Tsukuba, Ibaraki 305-8501, Japan}
\author{A. N. Cleland}
\affiliation{Department of Physics, University of California at Santa Barbara, Broida Hall, Santa Barbara, CA 93106}
\author{John M. Martinis}
\email{martinis@physics.ucsb.edu}
\affiliation{Department of Physics, University of California at Santa Barbara, Broida Hall, Santa Barbara, CA 93106}


\date{\today}
\begin{abstract}
Minimizing phase and other errors in experimental quantum gates allows higher fidelity quantum processing. To quantify and correct for phase errors in particular, we have developed a new experimental metrology --- amplified phase error (APE) pulses --- that amplifies and helps identify phase errors in general multi-level qubit architectures. In order to correct for both phase and amplitude errors specific to virtual transitions and leakage outside of the qubit manifold, we implement ``half derivative" an experimental simplification of derivative reduction by adiabatic gate (DRAG) control theory. The phase errors are lowered by about a factor of five using this method to $\sim 1.6^{\circ}$ per gate, and can be tuned to zero. Leakage outside the qubit manifold, to the qubit $|2\rangle$ state, is also reduced to $\sim 10^{-4}$ for $20\%$ faster gates.
\end{abstract}

\maketitle
Many candidate systems for quantum computation display several quantum energy levels, with computing architectures either employing just two of these levels\,\cite{Nielsen2000, Neeley2009, Cirac1995, Gershenfeld1997, Bianchetti2010, DiCarlo2009, You2005, Weber2010, Imamoglu1999, Platzman1999, Petta2005, Martinis2002}, or using three or more in (``\textit{d}-level") \textit{qudit}-based approaches\,\cite{Neeley2009, DiCarlo2009, Yamamoto2010}. Achieving good control over the computational states, while avoiding interference from the non-computational ones, requires measuring and understanding the source and magnitude of error-generating processes as well as correcting for such errors. For example, gate operations in the superconducting phase\,\cite{Neeley2009} and transmon\,\cite{DiCarlo2009} qubits, which are relatively weakly anharmonic, can generate leakage from the computational manifold to higher (non-computational) levels (``manifold leakage''). However, by careful experimental design one can identify and minimize the source of such errors, thereby improving qubit control.

Previous experiments identified the amplitude errors associated with manifold leakage by using error budgeting and metrology\,\cite{Lucero2008}. By directly measuring leakage using the Ramsey error filter protocol, this error is now understood and can be suppressed to $\sim 10^{-4}$\,\cite{Lucero2008}, at or near the threshold for fault-tolerant quantum computing\,\cite{Motzoi2009, Knill2005}. Randomized benchmarking, an alternative approach to measuring gate error, relies on a many-pulse protocol, effectively averaging over the Bloch sphere to quantify gate fidelity\,\cite{Knill2005, Chow2009}. Unfortunately, this approach optimizes a single value (gate fidelity) and does not distinguish between amplitude and phase errors. Quantum process tomography by contrast provides a complete analysis of gate operation\,\cite{Neeley2008, Bialczak2010, Nielsen2000}, but requires calibrated $X_{\pi/2}$ and $Y_{\pi/2}$ pulses, which can themselves be error sources.

Phase errors also contribute to overall gate error, but the relative contribution differs from amplitude-related errors.  Consequently, a method is needed to separately quantify the phase error generated by a gate. This will aid in identifying the source of these errors, improve calibration of high-fidelity tomography pulses, and provide a benchmark for optimizing control pulses.

In this Letter we explicitly show that gate errors can be separated into amplitude and phase errors. Virtual excitations to non-computational levels (e.g. the $\ket{2}$ state) during gate operation result in phase errors, while real excitations after the gate produce amplitude errors.  In order to better quantify the phase errors, we introduce a new metrology, ``amplified phase error" (APE) pulses, which uses a Ramsey fringe experiment to measure and amplify this ubiquitous source of error. We focus on errors related to $\pi/2$ pulses, because such pulses provide the basis for tomography and are essential in algorithms. We also demonstrate a simplified experimental version of the protocol\,\cite{Chow2010} termed ``derivative removal by adiabatic gates'' (DRAG)\,\cite{Motzoi2009}, which we call ``half-derivative'' (HD) pulses. By using HD pulses together with APE metrology, we measure and reduce the phase error to $1.6^{\circ}$ per gate, a factor of five reduction from unoptimized performance.  As a demonstration of this method, we perform quantum state tomography to map out the trajectories of typical HD pulses, including a $\pi$-pulse and a rotation about an arbitrary axis, using $X,Y$ and $Z$ controls to implement an (off-equator) Hadamard gate.  In addition, we show that APE metrology is a universal tool for probing phase errors on any of the higher qudit levels.

\begin{figure}[t!]
\includegraphics[trim = 0 0.1in 0 0.1in]{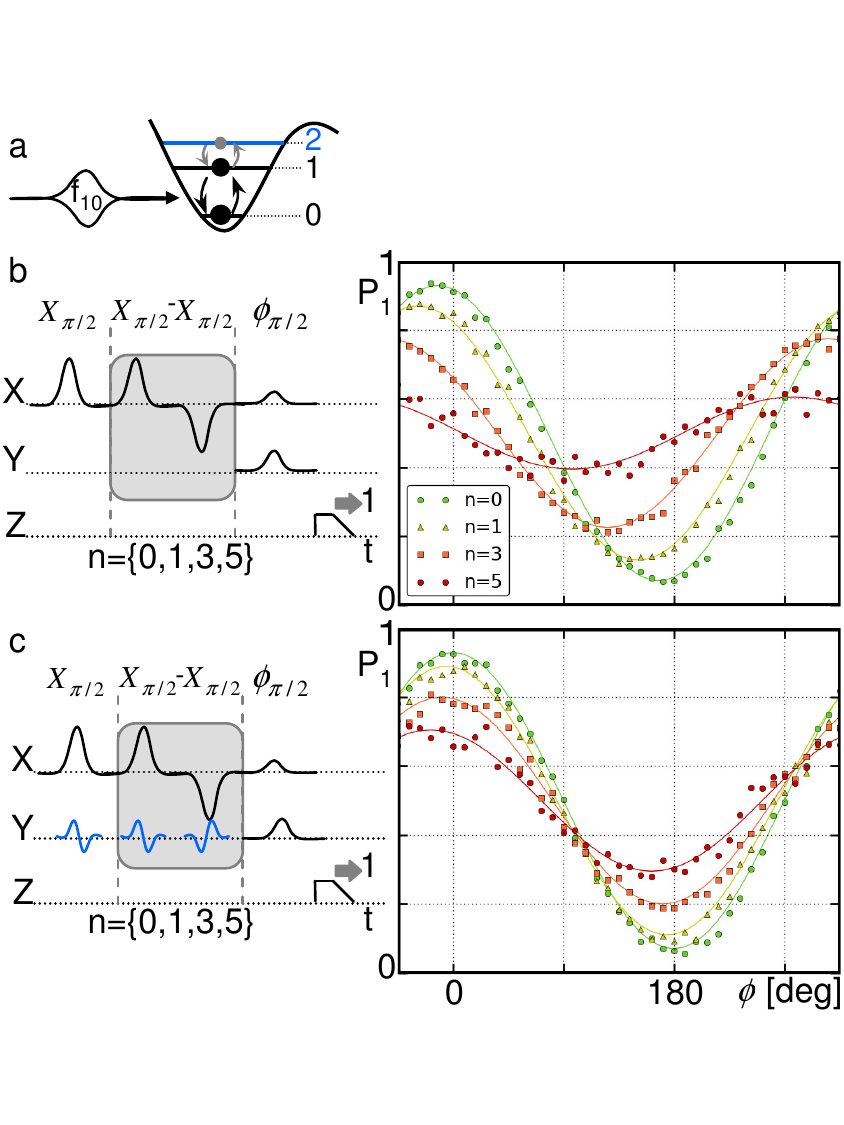}
\caption{(Color online) Multilevel qubit and amplified phase error (APE) metrology. ({\bf a}) Qubit with three energy levels, illustrating phase error due to virtual transitions to the $\ket{2}$ state. While performing on-resonance $\ket{0}\leftrightarrow\ket{1}$ gate operations at frequency $f_{10}$, virtual transitions to $\ket{2}$ create a phase change in $\ket{1}$. ({\bf b}) Left, single-control ($X$-quadrature) APE pulse sequence, where the pulse shape represents the microwave envelope. Ramsey-fringe experiments are modified by $I^{\prime n}_{\pi/2}$ ($n \in \{0,1,3,5\}$) pseudo-identity operations in-between the first $X_{\pi/2}$ and last $\phi_{\pi / 2}$ pulses. At the end the $Z$ control line is pulsed to measure the probability of $\ket{1}$. Right, the probability of measuring $\ket{1}$ as a function of the rotation axis $\phi$ of the final $\pi / 2$ pulse. Fits to extract the phase shift are plotted as lines with the data (dots). ({\bf c}) Left, same pulse sequence as in {\bf a}, but with the addition of $Y$-quadrature ``half-derivative" pulses, as discussed in the text. Right, the data (dots) with fits (lines) show small phase shifts for the HD pulse sequence.}
\label{fig:ape}
\end{figure}

In the experiments described here, we used a single superconducting phase qubit with $T_{1}=450\,\rm{ns}$ and $T^{\rm{echo}}_{2}=390\,\rm{ns}$. The circuit layout and operation have been described previously\,\cite{Neeley2009}. We have three-axis control over the qubit: microwave pulses of arbitrary amplitude and phase, resonant with the qubit $\ket{1} \leftrightarrow \ket{0}$ transition frequency $f_{10}$, produce rotations about any axis in the $x-y$ plane, while current pulses on the qubit bias line adiabatically change the qubit frequency, causing phase accumulation between $\ket{0}$ and $\ket{1}$, generating $z$-axis rotations\,\cite{Steffen2006a}. The $\ket{2} \leftrightarrow \ket{1}$ transition frequency $f_{21}$ differs from $f_{10}$ by $\Delta/2\pi = f_{21} - f_{10} \sim -200\ \rm{MHz}$ (see Fig. \ref{fig:ape}a).

Non-ideal qubit behavior can arise from both leakage at the end of the gate and virtual transitions to higher states during on-resonant operations. The leakage is an amplitude error, representing loss of probability to states outside the manifold. Leakage can be reduced to $\sim 10^{-4}$ by careful shaping of the microwave envelope and choosing the correct gate duration, which scales as $1/\Delta$\,\cite{Lucero2008, Steffen2003}.

The phase error arising from virtual transitions (especially to the $\ket{2}$ state) can be modeled as effective qubit rotations about the $z$-axis. We first restrict ourselves to simple gates comprising $\pi$ and $\pi/2$ rotations. An $X_{\pi/2}$ pulse (a rotation about the $x$-axis by an angle $\theta = \pi/2$) ideally produces the transformation
\begin{eqnarray}
X_{\pi/2}   =  e^{-i\sigma_{x}\frac{\pi}{4}}
 =  \frac{1}{\sqrt{2}}
	\begin{pmatrix}
	1 & -i \\
	-i & 1
	\end{pmatrix}\ ,
\end{eqnarray}
where $\sigma_{x}$ is one of the Pauli matrices. From numerical simulations of our multi-level qubit, we find that this transformation, expressed in quantum circuit language, is instead
\begin{eqnarray}
X^{\prime}_{\pi/2}=e^{-i\epsilon^{\prime}}Z_{\epsilon}X_{\pi/2}Z_{\epsilon}\ ,
\label{eq:xep}
\end{eqnarray}
where $Z_{\epsilon}$ is the phase error of interest and $0<\epsilon\ll1$\,\cite{supp}. The leading term in Eq.\,(\ref{eq:xep}) is  a global phase and can be ignored. We note that Eq.\,(\ref{eq:xep}) differs from $X^{\star}_{\pi/2}=Z_{-\epsilon}X_{\pi/2}Z_{\epsilon}$, which corresponds to a rotation about a new axis $\epsilon$ away from the $x$-axis in the $x-y$ plane.

The phase error $\epsilon$ is a function of both the rotation angle $\theta$ and the gate time $t_{g}$. From simulations, we find that $\epsilon \sim \theta^{2}/t_{g}$. Longer gate times decrease the virtual transitions and consequently reduce the phase error, consistent with the AC-Stark effect\,\cite{Shore1978}.

In order to best reduce this error, we first sought a protocol that would amplify the error $\epsilon$. If we consider a $2\pi$ rotation generated by concatenating four $\pi/2$ pulses, we find from Eq.\,(\ref{eq:xep}) that $X^{\prime 4}_{\pi/2}=-e^{-4i\epsilon^{\prime}}I$, where $I$ is the identity\,\cite{supp}. A concatenated $2\pi$ rotation thus does not accumulate the relative phase error.

We next examine the pseudo-identity operation that is formed by concatenating positive and negative $\theta$ rotations. For a first-order expansion with $\epsilon \ll 1$ we find
\begin{eqnarray}
I^{\prime}_{\theta} & = & (Z_{\epsilon}X_{\theta}Z_{\epsilon})(Z_{\epsilon}X_{-\theta}Z_{\epsilon})  \nonumber \\
	& \approx &
	 \begin{pmatrix}
	1+i(\cos\theta-1)\epsilon & -(\sin\theta)\epsilon \\
	(\sin\theta)\epsilon & 1-i(\cos\theta +3)\epsilon \\
	\end{pmatrix}\ ,
\end{eqnarray}
where $X_{\theta}$ is an arbitrary rotation of $\theta$ about the $x$-axis\,\cite{supp}. For $\theta = \pi$ we find that $I^{\prime}_{\pi}=e^{-2i\epsilon^{\prime}}I$, which is similar to the $2\pi$ rotation, as the phase error $\epsilon$ cancels. However, for $\theta = \pi/2$ we find
\begin{eqnarray}
I^{\prime}_{\pi/2} & = & (Z_{\epsilon}X_{-\pi/2}Z_{\epsilon})(Z_{\epsilon}X_{\pi/2}Z_{\epsilon}) \nonumber \\
	& \approx &
	 \begin{pmatrix}
	1-i\epsilon & \epsilon \\
	-\epsilon & 1-3i\epsilon \\
	\end{pmatrix}\ ,
\label{eq:identity}
\end{eqnarray}
showing phase error accumulation.

To measure this error, we combine the result from Eq.\,(\ref{eq:identity}) with a Ramsey fringe experiment, forming the ``amplified phase error" (APE) sequence. The APE sequence consists of inserting $n$ successive $I^{\prime}_{\pi/2}$ pseudo-identity operations between the $\pi/2$ pulses that define a Ramsey fringe measurement (Fig. \ref{fig:ape}b). The phase error is amplified by $2 n$ for $n$ applications of the pseudo-identity operation,
\begin{eqnarray}
I^{\prime n}_{\pi/2}  \approx (Z_{2\epsilon})^{n} = Z_{2 n \epsilon}\ .
\label{eq:iz}
\end{eqnarray}
By applying APE pulses to the state $\ket{\psi}=(\ket{0}-i\ket{1})/\sqrt{2}$ followed by a final $\phi_{\pi/2}$ pulse, we directly probe the phase error due to the $X_{\pi/2}$ pulses.

Figure\,\ref{fig:ape}b shows the probability of measuring the $\ket{1}$ state versus rotation axis $\phi$ of the final $\phi_{\pi/2}$ pulse, for $I^{\prime n}_{\pi/2}$ ($n=0,1,3,5$) pseudo-identity operations. Consistent with Eq.\,(\ref{eq:iz}), the phase error scales with $n$. For $n=5$ in Fig.\,\ref{fig:ape}b, the final pulse is $\sim 90^{\circ}$ out of phase, corresponding to a $10 \times$ phase error amplification from a total of $11$ pulses, yielding $7.3^{\circ}$ phase error per gate. The oscillation amplitude is also reduced, due to decoherence.

\begin{figure}[tb!]
\includegraphics[width=3.3in, trim = 0 0.8in 0 0.8in, clip]{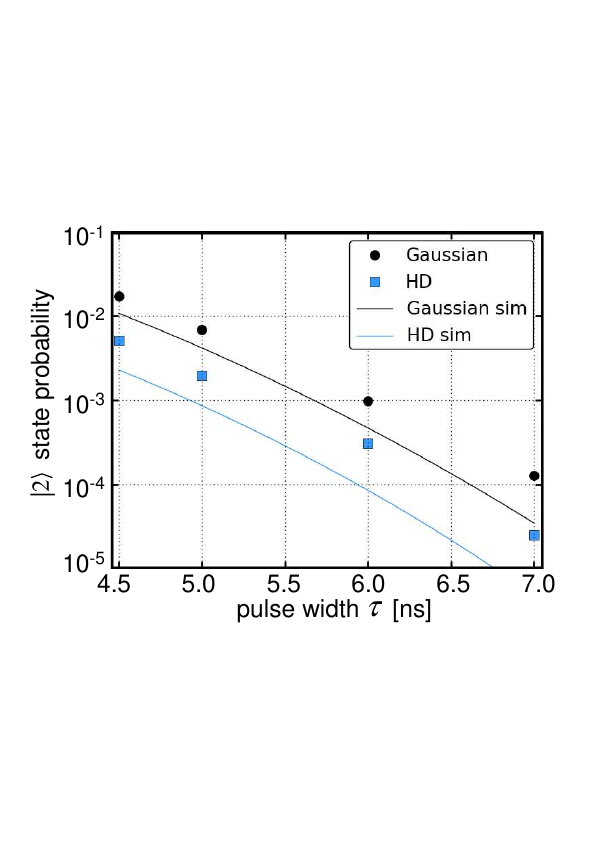}
\caption{(Color online) Amplitude errors due to leakage into the $\ket{2}$ state from an $X_{\pi}$-pulse. Plot of $\ket{2}$ error versus pulse width (FWHM) $\tau$ for single-control Gaussian (black dots) and HD (blue squares) $\pi$-pulses. The $6\,\rm{ns}$ HD pulse produces $\times 5$ lower error and $20\%$ faster gates. The solid lines are three-state simulations using Gaussian (black) and HD (blue) pulses.}
\label{fig:ramsey}
\end{figure}

To correct the phase error, we employ the derivative reduction by adiabatic gates (DRAG) protocol\,\cite{Motzoi2009}. The original DRAG prescription uses three controls, $X$, $Y$, and $Z$. The $X$ control provides the original envelope-shaping to the microwaves, which we implemented as a Gaussian in time with arbitrary amplitude $A$, $X = A \,\exp[-4\ln(2)(t-t_{0})^2/\tau^2]$, where $\tau$ is the full-width-at-half-maximum (FWHM) and $t_{0}$ the time at the center of the pulse. The quadrature control $Y=-\dot{X}/{\Delta}$ is the time derivative of the $X$ control scaled by the nonlinearity $\Delta$. The $Z$ control produces a dynamic detuning pulse during the gate that removes the effective $z$-rotations from the virtual transitions.

We find both in simulations and experiment that the $Y$ and $Z$ controls are not independent. For experimental simplicity,  we set $Z$ to zero and compensate by reducing the magnitude of the $Y$ control by $1/2$, to form the so-called ``half-derivative" (HD) protocol\,\cite{supp}. For a Gaussian envelope on the $X$ control, the HD pulses are as illustrated in Fig.\,\ref{fig:ape}c and differ from the DRAG pulses by the quadrature controls, $Y= -\dot{X}/(2\Delta)$, $Z = 0$. Experimentally, we found that HD performed as well as full DRAG.

The HD pulse sequence in Fig.\,\ref{fig:ape}c is the same as Fig.\,\ref{fig:ape}b, with the addition of the $Y$ control. Data are plotted for the same number of $I^{\prime}_{\pi/2}$ pseudo-identity operations. We find by applying the HD protocol the phase error is reduced to $1.6^{\circ}$ per gate, and can be further minimized by tuning the amplitude of the derivative pulse.

\begin{figure}[t!]
\includegraphics[width=3.1in, trim = 0 0.0in 0 0.0in, clip]{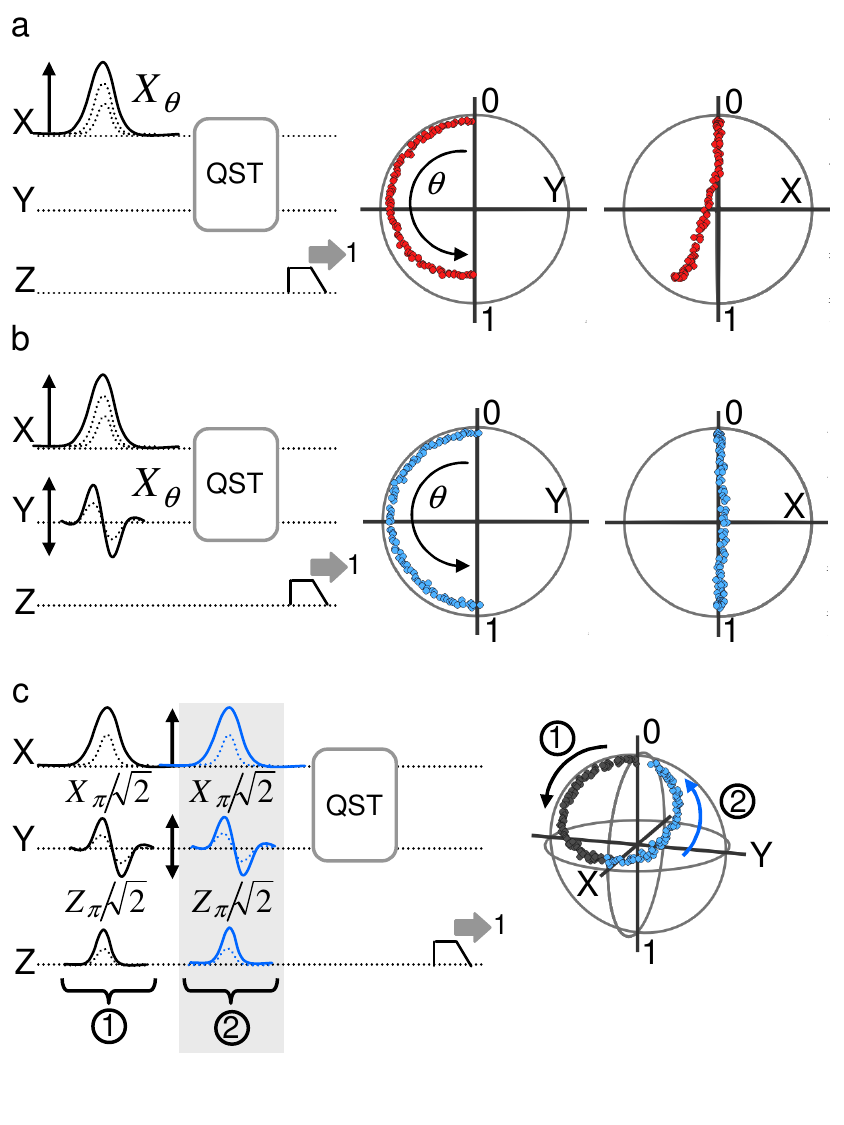}
\caption{(Color online) Gate trajectories using quantum state tomography (QST) for single control and HD pulses. ({\bf a}) Left, pulse sequence for an $X$ rotation. A Gaussian pulse of fixed length ($\rm{FWHM}=6\,\rm{ns}$) and varying amplitude completes the rotation angle to $\theta = \pi$. QST is performed at every incremental increase of amplitude. Right, the reconstructed quantum state data is plotted on the Bloch sphere from two perspectives, looking down the $x$ and the $-y$ axes. ({\bf b}) Same as in {\bf a} except using HD ($X$ and $Y$ simultaneous control) pulses to perform the $X$ rotation. No phase error is observed. ({\bf c}) Left, pulse sequence of a two-part trajectory using HD pulses and $Z$ control to form a Hadamard gate. $X$, $Y$, and $Z$ controls are fixed-length ($\rm{FWHM}=6\,\rm{ns}$) pulses with increasing amplitudes to execute $\pi/\sqrt{2}$ rotations about $X$ and $Z$. Trajectory 2 pulses ramp-up only after trajectory 1 pulses are at full amplitude. Right, the reconstructed QST data. Each trajectory completes a Hadamard gate, taking $\ket{0} \rightarrow (\ket{0}+\ket{1})/\sqrt{2} \rightarrow \ket{0}$.}
\label{fig:traj}
\end{figure}

HD pulses also reduce the leakage to the $\ket{2}$ state. Plotted in Fig.\,\ref{fig:ramsey} are the data from a Ramsey error filter\,\cite{Lucero2008} for both single control Gaussian and HD pulses. A $6\,\rm{ns}$ (FWHM) HD $X_{\pi}$ pulse gives a $\ket{2}$ state probability of $10^{-4}$, almost an order of magnitude better than a non-HD pulse of the same width, which consequently provides a $20\%$ faster gate\,\cite{Lucero2008}.

With calibrated $X_{\pi/2}$ and $Y_{\pi/2}$ pulses, we can perform quantum state tomography (QST) without worry of miscalibrated measurement axes. As a practical demonstration of how HD pulses reduce phase error, we perform QST\,\cite{Steffen2006a} with and without HD. Figure \ref{fig:traj}a(b) shows the pulse sequence and data for the Gaussian pulses (HD pulses) during an $X_{\theta}$ rotation. The pulses are of fixed length ($\rm{FWHM}=6\,\rm{ns}$) with variable amplitude $\theta$. QST is performed at each incremental increase of amplitude and the quantum state is recreated in the Bloch sphere as shown to the right of each of the respective pulse sequences. In contrast with the single control Gaussian pulses, the HD pulses execute a meridian trajectory with no phase error with increasing $\theta$.

The final HD demonstration is an (off-equator) Hadamard gate, shown in Fig.\,\ref{fig:traj}c, which uses all three control lines\,\cite{supp}. We incrementally increase the amplitude of all three control lines using fixed length ($\rm{FWHM}=6\,\rm{ns}$) pulses to perform rotations from $0$ to $\pi/\sqrt{2}$ about both the $x$ and $z$ axes, which at full amplitude gives the Hadamard gate $H$ ($\ket{0} \rightarrow ( \ket{0} + \ket{1})/\sqrt{2}$). The trajectory concludes with a second set of pulses to complete the identity operation $I=HH$, and returning to the initial state $( \ket{0} + \ket{1})/\sqrt{2} \rightarrow \ket{0}$.

We also consider the challenge of optimizing control pulses for each $d$-level of a qudit\,\cite{Neeley2009}. Tomography verifies the operation, but again relies on the calibrations of $\pi/2$ pulses. In principle, APE provides the necessary phase calibration certification for tomography.

\begin{figure}[!t]
\includegraphics[width=3.39in, trim = 0 1.4in 0 0.3in, clip]{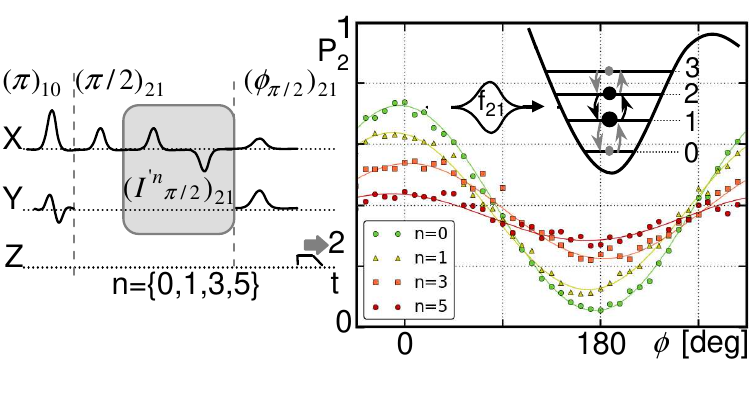}
\caption{(Color online) Multilevel qudit and APE metrology for the $\ket{1} \leftrightarrow \ket{2}$ transition. Inset, the four energy levels of the qudit illustrate the nearly symmetric virtual transitions to states $\ket{3}$ and $\ket{0}$ that mostly cancel the phase error during the $f_{21}$ resonant drive. Left, pulse sequence showing $(\pi)_{10}$ for a $\ket{0} \rightarrow \ket{1}$ transition then a Ramsey fringe with the pseudo-identity operation on the $\ket{1} \leftrightarrow \ket{2}$ transition. Right, the probability of measuring the $\ket{2}$ state as a function of the rotation axis of the final $(\phi_{\pi/2})_{21}$ pulse for $n \in \{0,1,3,5\}$ pseudo-identity operations. With only a single control pulse (no HD correction), the phase error is $1.1^{\circ}$ per gate.}
\label{fig:genape}
\end{figure}

To demonstrate the general utility of the APE protocol, we implement this metrology on the qudit level for state $\ket{2}$. We first calibrate the $(\pi)_{10}$ and $(\pi)_{21}$ pulses to generate the $\ket{0}\rightarrow\ket{1}$ and $\ket{1}\rightarrow\ket{2}$ transitions, respectively\,\cite{Neeley2009}. As shown in Fig.\,\ref{fig:genape}, we first prepare the $\ket{1}$ state via an HD $\pi$ pulse so that we can then perform a Ramsey fringe experiment using the $\ket{1}$ and $\ket{2}$ states. The APE pulses are applied between the first and last $(\pi/2)_{21}$ pulses, only now resonant with $f_{21}$. The data are for single controlled Gaussian envelope pulses, i.e. no HD protocol for the $\ket{1} \rightarrow \ket{2}$ pulses.

Surprisingly, after $n=5$ pseudo-identity operations, only $12^{\circ}$ of phase error is measured, equivalent to $1.1^{\circ}$ per gate. We offer a qualitative interpretation: the relatively small anharmonicity of the phase qudit, combined with the symmetric virtual transitions to the $\ket{3}$ and $\ket{0}$ states, provide complementary phase shifts that partially cancel out the phase error.

In conclusion, we introduced a new metrology tool, amplified phase error (APE) pulses, which can amplify the phase error by an order of magnitude. Together with APE and half-derivative pulses, our simplified variant of DRAG\,\cite{Motzoi2009}, we identify and reduce phase errors to $1.6^{\circ}$ per gate. By simply re-scaling the analytic form for the HD pulses, the phase error can be completely removed. The HD pulses also can increase gate speed by $20\%$.

\begin{acknowledgments}
We acknowledge discussions with Jerry Chow, Jay Gambetta, Felix Motzoi, and Frank Wilhelm. Devices were made at the UCSB Nanofabrication Facility, part of the NSF-funded National Nanotechnology Infrastructure Network. M.M. acknowledges support from an Elings fellowship. This work was supported by IARPA under grant W911NF-04-1-02-4.
\end{acknowledgments}

\bibliographystyle{apsrev}

\begin{thebibliography}{3}
\email{martinis@physics.ucsb.edu}

\bibitem{Nielsen2000} M.~A. Nielsen, I.~L. Chuang \textit{Quantum Computation and Quantum Information} (Cambridge Univ. Press, Cambridge 2000).

\bibitem{Neeley2009} M. Neeley \textit{et al.}, Science \textbf{325}, 722 (2009).

\bibitem{Cirac1995} J.~I. Cirac and P. Zoller, Phys. Rev. Lett \textbf{74}, 4091 (1995).

\bibitem{Gershenfeld1997} N.~A. Gershenfeld and I.~L. Chuang, Science \textbf{275}, 350 (1997).

\bibitem{Bianchetti2010} R. Bianchetti \textit{et al.}, arXiv:1004.5504v1 (2010).

\bibitem{DiCarlo2009} L.DiCarlo \textit{et al.}, Nature \textbf{460}, 240 (2009).

\bibitem{You2005} J.~Q. You, F. Nori, Phys. Today \textbf{58}, 42--47 (2005).

\bibitem{Imamoglu1999} A. Imamoglu \textit{et al.}, Phys. Rev. Lett. \textbf{83}, 4202 (1999)

\bibitem{Weber2010} J.~R. Weber \textit{et al.}, Proc. Natl. Acad. Sci. \textbf{107}, 8513 (2010)

\bibitem{Platzman1999} P.~M. Platzman and M.~I. Dykman, Science \textbf{284}, 1967 (1999)

\bibitem{Petta2005} J.~R. Petta \textit{et al.}, Science \textbf{309}, 2180�2184 (2005)

\bibitem{Martinis2002} J.~M. Martinis \textit{et al.}, Phys. Rev. Lett. \textbf{89}, 117901 (2002).

\bibitem{Yamamoto2010} T. Yamamoto  \textit{et al.}, arXiv:1006.5084v1.

\bibitem{Lucero2008} E. Lucero \textit{et al.},  Phys. Rev. Lett. \textbf{100}, 247001 (2008).

\bibitem{Motzoi2009} F. Motzoi, \textit{et al.}, Phys. Rev. Lett. \textbf{103}, 110501 (2009).

\bibitem{Knill2005} E. Knill, Nature \textbf{434}, 39--44 (2005).

\bibitem{Chow2009} J.~M. Chow \textit{et al.},  Phys. Rev. Lett. \textbf{102}, 090502 (2009).

\bibitem{Neeley2008} M. Neeley \textit{et al.}, Nature Physics \textbf{4}, 523 (2008).

\bibitem{Bialczak2010} R.~C. Bialczak \textit{et al.}, Nature Physics \textbf{6}, 409 (2010).

\bibitem{Chow2010}  J.~M. Chow \textit{et al.},  arXiv:100501v1.

\bibitem{Steffen2006a} M. Steffen \textit{et al.}, Phys. Rev. Lett. \textbf{97}, 050502 (2006).

\bibitem{Steffen2003} M. Steffen, J.~M. Martinis, I.~L. Chuang, Phys. Rev. B \textbf{68}, 224518 (2003).

\bibitem{supp} See supplemental material for more details.

\bibitem{Shore1978} B.~W. Shore Phys. Rev. A \textbf{17}, 1739 (1978).

\end{thebibliography}

\end{document}

\title{Supplemental: Reduced phase error through optimized control \\ of a superconducting qubit}
\author{Erik Lucero}
\affiliation{Department of Physics, University of California at Santa Barbara, Broida Hall, Santa Barbara, CA 93106}
\author{Julian Kelly}
\affiliation{Department of Physics, University of California at Santa Barbara, Broida Hall, Santa Barbara, CA 93106}
\author{Radoslaw C. Bialczak}
\affiliation{Department of Physics, University of California at Santa Barbara, Broida Hall, Santa Barbara, CA 93106}
\author{Mike Lenander}
\affiliation{Department of Physics, University of California at Santa Barbara, Broida Hall, Santa Barbara, CA 93106}
\author{Matteo Mariantoni}
\affiliation{Department of Physics, University of California at Santa Barbara, Broida Hall, Santa Barbara, CA 93106}
\author{Matthew Neeley}
\affiliation{Department of Physics, University of California at Santa Barbara, Broida Hall, Santa Barbara, CA 93106}
\author{A. D. O'Connell}
\affiliation{Department of Physics, University of California at Santa Barbara, Broida Hall, Santa Barbara, CA 93106}
\author{Daniel Sank}
\affiliation{Department of Physics, University of California at Santa Barbara, Broida Hall, Santa Barbara, CA 93106}
\author{H. Wang}
\affiliation{Department of Physics, University of California at Santa Barbara, Broida Hall, Santa Barbara, CA 93106}
\author{Martin Weides}
\affiliation{Department of Physics, University of California at Santa Barbara, Broida Hall, Santa Barbara, CA 93106}
\author{James Wenner}
\affiliation{Department of Physics, University of California at Santa Barbara, Broida Hall, Santa Barbara, CA 93106}
\author{Tsuyoshi Yamamoto}
\affiliation{Department of Physics, University of California at Santa Barbara, Broida Hall, Santa Barbara, CA 93106}
\affiliation{Green Innovation Research Laboratories, NEC Corporation, Tsukuba, Ibaraki 305-8501, Japan}
\author{A. N. Cleland}
\affiliation{Department of Physics, University of California at Santa Barbara, Broida Hall, Santa Barbara, CA 93106}
\author{John M. Martinis}
\email{martinis@physics.ucsb.edu}
\affiliation{Department of Physics, University of California at Santa Barbara, Broida Hall, Santa Barbara, CA 93106}

\pacs{03.67.Ac, 03.67.Lx, 03.67.Pp, 74.50+r, 78.47.jm, 85.25.Cp}
\keywords{Josephson Junction, Quantum Computing, phase error, High Fidelity Gates}

\date{\today}
\maketitle
\section{Tracking qubit frequency}
Optimized qubit control pulses rely on a precise measurement of the qubit transition frequency. After performing spectroscopy to find  the qubit frequency $f_{10}$ to within $\sim1\,\rm{MHz}$, we fix the gate time $\tau$ and tune the microwave amplitude to execute a $\pi$ rotation. We verify that the computed amplitude for $\pi/2$ rotation is indeed half that for a $\pi$ rotation by performing two consecutive $\pi/2$ rotations and comparing the probability $P_{1}$ between that of a single $\pi$ rotation. 

Next, to precisely measure the qubit frequency, we use a Ramsey fringe experiment, where the final $\phi_{\pi/2}$-pulse rotates at $50 \rm{MHz}$ about a variable axis on the equator of the Bloch sphere. A frequency shift in the oscillations of $P_{1}$ different from $50\rm{MHz}$ is the amount the microwave drive is detuned from the qubit frequency. Correcting for this offset precisely tunes the microwave drive to the qubit frequency to within $1$ part in $10^4$ (sub $\rm{MHz}$ resolution), which is consistent with limits set from 1/f flux-noise fluctuations\cite{Bialczack2007}. 

\begin{figure}[h!]
\includegraphics[width=3.27in, trim = 0 0.6in 0 0.0in, clip]{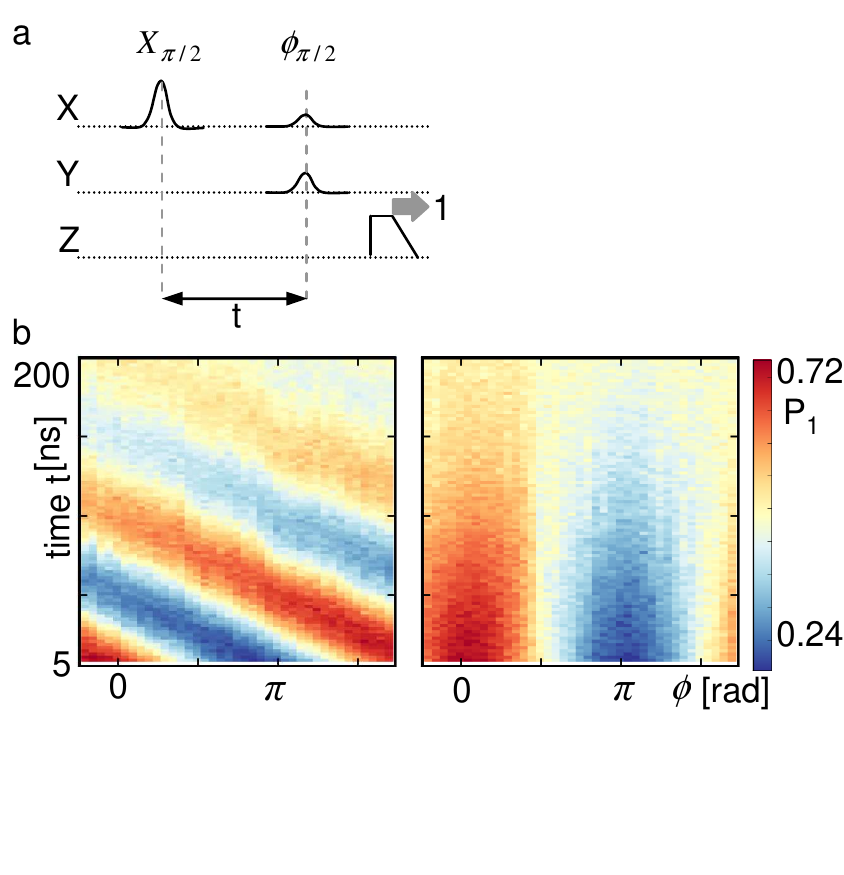}
\caption{(Color online)Tracking the qubit frequency $f_{10}$. ({\bf a}) Ramsey fringe sequence that consists of two Gaussian-shaped $\pi/2$-pulses, separated in time by $t$, followed by a measure pulse on the Z-control tuned to tunnel the $\ket{1}$ state. The first $\pi/2$-pulse defines the rotation axis; by convention this is the x-axis. The second pulse is delayed by a time $t$ with variable rotation axis $\phi$. ({\bf b}) Plots of $P_{1}$ versus separation time $t$ and phase $\phi$ for pulse sequence in {\bf a}. Left, microwave drive is $10\,\rm{MHz}$ detuned from the qubit frequency $f_{10}$. Right, data taken after calibration; microwaves detuned less than $1\,\rm{MHz}$ from the qubit frequency.}
\label{fig:freqtuner}
\end{figure}

We also can perform a more complete test (2-D scans) of this frequency calibration by noting that the phase error is $\epsilon=\delta f \delta t$ if the microwave drive is $\delta f$ off-resonance from the qubit frequency for some time $\delta t$. Therefore, we verify that the microwave carrier matches the qubit frequency for the entire $\delta t$ of the APE sequences via a Ramsey fringe experiment as shown in Fig.\,\ref{fig:freqtuner}. Fig.\,\ref{fig:freqtuner}b shows the data for two different detunings: left,  $\delta f = 10\,\rm{MHz}$ and right, $\delta f <1\,\rm{MHz}$ (after performing the calibration described above). When the microwave drive is detuned by $\delta = 10\,\rm{MHz}$ the data show a distinct tilt and clear oscillations with a frequency of $10\,\rm{MHz}$. After calibration the data has no beating and therefore no sign of a detuned microwave drive. This confirms that the qubit frequency is tracked precisely throughout the duration of the desired (Amplified Phase Error) sequences.

To compute the nonlinearity $\Delta/2\pi = f_{21}-f_{10}$ for use in the HD protocol, we directly measure the transition frequency between states $\ket{1}$ and $\ket{2}$ with a Ramsey Error Filter (REF)\cite{Lucero2008}. The REF uses an oscillation, provides finer resolution, and is simpler to automate experimentally than for peak-finding in high power spectroscopy. To increase the $\ket{2}$ state population, the $(\pi)_{10}$ pulses (for the $\ket{0} \leftrightarrow \ket{1}$ transitions) are sufficiently short and do not use the HD protocol. Using this technique, we measure the $\ket{1} \rightarrow \ket{2}$ transition frequency, $f_{21}$ to within $1\,\rm{MHz}$.

\section{Amplified Phase Error Theory}
\begin{figure}[tb!]
\includegraphics[width=3.28in, trim = 0 0.9in 0 0.01in, clip]{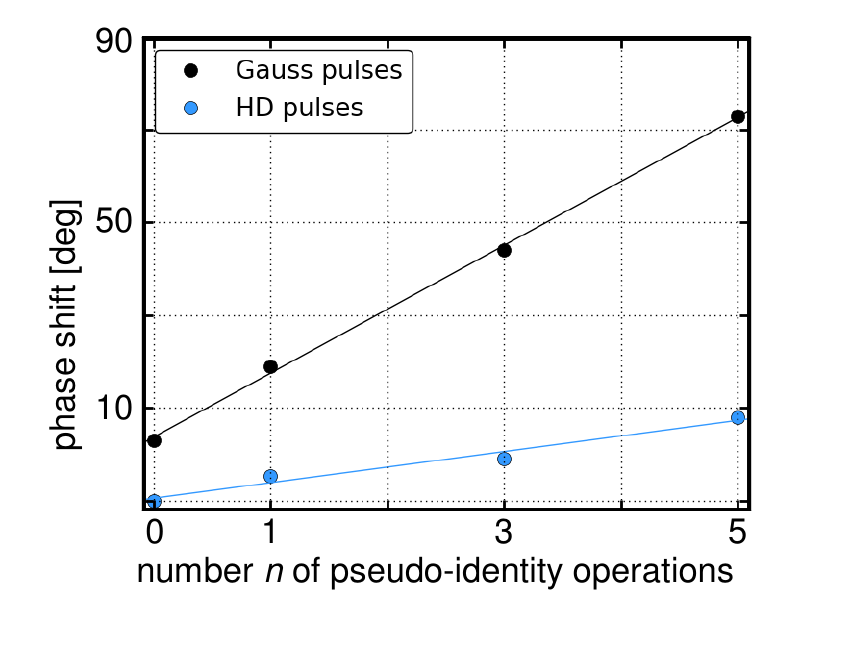}
\caption{Phase shift versus number \textit{n} of pseudo-identity operations.}
\label{fig:nident}
\end{figure}
From numerical simulations, an $X_{\pi/2}$ pulse has the transformation, (ignoring global phases), given by
\begin{eqnarray}
X^{\prime}_{\pi/2} \simeq Z_{\epsilon}X_{\pi/2}Z_{\epsilon}\, ,
\label{eq:xep}
\end{eqnarray}
where $Z_{\epsilon}$ is the phase error of interest
\begin{eqnarray}
Z_{\epsilon} & = &
\begin{pmatrix}
1 & 0 \\
0 & e^{-i\epsilon} \\
\end{pmatrix}\, .
\label{eq:zerr}
\end{eqnarray}
Using Eqs.\,(\ref{eq:xep}) and (\ref{eq:zerr}), we explicitly calculate the phase shift for a $2\pi$ rotation coming from four $\pi/2$ pulses. For an arbitrary rotation $\theta$ about the $x$-axis, the gate operation is
\begin{eqnarray}
X_{\theta} =
\begin{pmatrix}
\cos\theta/2 & -i\sin\theta/2 \\
-i\sin\theta/2 & \cos\theta/2 \\ 
\end{pmatrix} \ ,
\end{eqnarray}
such that $X^{\prime}_{\pi/2}$ is
\begin{eqnarray}
X^{\prime}_{\pi/2} & = & \frac{1}{\sqrt{2}}
\begin{pmatrix}
1 & -ie^{-i \epsilon} \\
-ie^{-i \epsilon} & e^{-i2\epsilon}  \\ 
\end{pmatrix}\, . \\
\end{eqnarray}

Concatenating four positive $\pi/2$ rotations results in
\begin{widetext}
\begin{eqnarray}
X^{\prime 4}_{\pi/2}  & \equiv & (X^{\prime}_{\pi/2})^{4} \nonumber \\
& = & \frac{1}{4}
\begin{pmatrix}
e^{-6i\epsilon}(-1 - e^{2i\epsilon} - 3e^{4i\epsilon} + e^{i\epsilon}) &  -ie^{-7i\epsilon}(-1 + e^{2i\epsilon})^2(1+e^{2i\epsilon}) \\
-ie^{-7i\epsilon}(-1 + e^{2i\epsilon})^{2} (1+e^{2i\epsilon}) & e^{-8i\epsilon}(-1 + 3e^{2i\epsilon} + e^{4i\epsilon} + e^{6i\epsilon})
\end{pmatrix} \nonumber \\
& \simeq & e^{-i4\epsilon}I \, ,
\label{eq:x4}
\end{eqnarray}
\end{widetext}
where $I$ is the identity. Equation\,(\ref{eq:x4}) only acquires a global phase.

Next, we calculate the phase shift for the pseudo-identity operation, used in the APE protocol, comprised of a positive then a negative $\theta=\pi/2$ rotation.
\begin{eqnarray}
I^{\prime}_{\pi/2} & =  & X^{\prime}_{-\pi/2} X^{\prime}_{\pi/2} \nonumber \\
  & = & 	
	 \begin{pmatrix}
	e^{-i \epsilon}\cos(\epsilon) & e^{-2 i \epsilon}\sin(\epsilon) \\
	-e^{-2 i \epsilon}\sin(\epsilon)  & e^{-3 i \epsilon}\cos(\epsilon) \\
	\end{pmatrix} \nonumber \\
	& \approx &
	 \begin{pmatrix}
	1-i\epsilon & \epsilon \\
	-\epsilon & 1-3 i\epsilon \\
	\end{pmatrix}\, .
\label{eq:identity}
\end{eqnarray}
For \textit{n} applications of the pseudo-identity operation, in the limit where $0 < \epsilon \ll 1$, $\epsilon \rightarrow n\epsilon$
\begin{eqnarray}
I^{\prime n}_{\pi/2} & \approx & 
 \begin{pmatrix}
	1-i n\epsilon & n\epsilon \\
	-n\epsilon & 1-3 i n\epsilon \\
	\end{pmatrix}\, ,
\end{eqnarray}
and by removing an overall global phase
\begin{eqnarray}	
I^{\prime n}_{\pi/2} & \approx & (Z_{2\epsilon})^{n} = Z_{2n\epsilon}\, .
\end{eqnarray}

The measured phase shift scales with the number n of pseudo-identity operations as shown in Fig.\,\ref{fig:nident}.

\section{Z-pulse calibration}

\begin{figure}[h!]
\includegraphics[width=3.28in, trim = 0 2.41in 0 0.0in, clip]{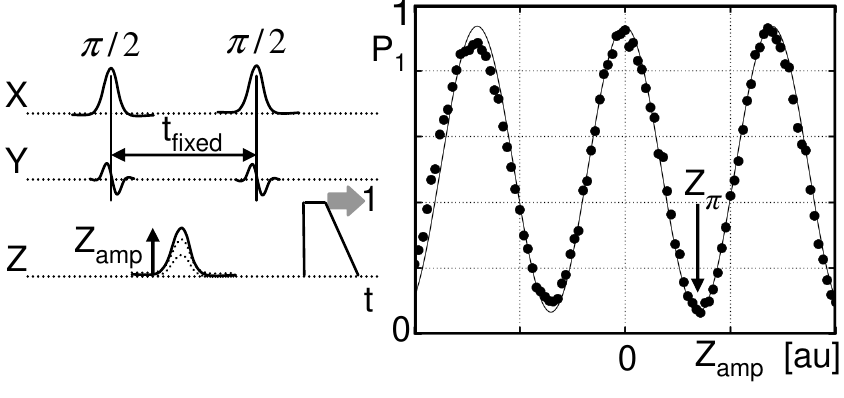}
\caption{$Z_{\pi}$ calibration. ({\bf a}) Left, the Ramsey-type pulse sequence to calibrate a $Z_{\pi}$ with the $X$ and $Y$ controls using the HD protocol described in the manuscript. The sequence consists of two, $6\rm{ns}$ (FWHM) $\pi/2$- pulses fixed in time with $t_{\rm{fixed}} = 24\rm{ns}$ and a $6\rm{ns}$ (FWHM) Z-pulse centered in between them. The separation time is chosen to minimize overlap of the pulses. The $Z_{\rm{amp}}$ increases incrementally. Right, the probability of measuring the $\ket{1}$ state $P_{1}$ as a function of Z-pulse amplitude, $Z_{\rm{amp}}$. The data are plotted as points with best fit as a line. $P_{1}$ oscillates with increasing magnitude of the Z-pulse amplitude. Arrow indicates the Z-pulse amplitude equivalent to a $\pi$ rotation about the Z-axis.}
\label{fig:ztuner}
\end{figure}

Explicit Z gates are required for the (off-equator) Hadamard gate. We calibrate our Z pulse as shown in Fig.\,\ref{fig:ztuner}. A static length (full-width at half-max $=6\rm{ns}$) with an increasing amplitude Z-pulse is inserted between two HD $\pi/2$ pulses with fixed separation time $t_{\rm{fixed}}$. The probability of measuring the $\ket{1}$ state $P_{1}$ oscillates with increasing $Z_{\rm{amp}}$\cite{Steffen2006a}. The arrow indicates the $Z_{\rm{amp}}$ that corresponds to a rotation angle of $\pi$ about the $z$-axis.

\bibliographystyle{apsrev}